\documentclass[emulateapj]{article}
\usepackage[onecolumn]{emulateapj}
\newcommand{\calP}{{\cal P}}

\def\kmsmpc{\,{\rm km\,s^{-1}\,Mpc^{-1}}}

\def\punits{\,{\rm cm^{-2}\,s^{-1}}}

\def\sfrd{\,{\rm M_\odot\,yr^{-1}\,Mpc^{-3}}}

\def\etal{{et al.\ }}

\def\spose#1{\hbox to 0pt{#1\hss}}
\def\lta{\mathrel{\spose{\lower 3pt\hbox{$\mathchar"218$}}
     \raise 2.0pt\hbox{$\mathchar"13C$}}}
\def\gta{\mathrel{\spose{\lower 3pt\hbox{$\mathchar"218$}}
     \raise 2.0pt\hbox{$\mathchar"13E$}}}

\newcommand{\mincir}{\raise -2.truept\hbox{\rlap{\hbox{$\sim$}}\raise5.truept
\hbox{$<$}\ }}
\newcommand{\magcir}{\raise -2.truept\hbox{\rla669p{\hbox{$\sim$}}\raise5.truept
\hbox{$>$}\ }}
\newcommand{\minmag}{\raise-2.truept\hbox{\rlap{\hbox{$<$}}\raise 6.truept\hbox
{$>$}\ }}

\newcommand{\be}{\begin{equation}}
\newcommand{\ee}{\end{equation}}
\newcommand{\ba}{\begin{eqnarray}}
\newcommand{\ea}{\end{eqnarray}}
\newcommand{\brr}{\begin{array}}
\newcommand{\err}{\end{array}}
\newcommand{\bc}{\begin{center}}
\newcommand{\ec}{\end{center}}
\newcommand{\f}{\frac}
\newcommand{\Om}{\Omega}
\newcommand{\de}{\delta}
\newcommand{\s}{\sigma}

\lefthead{Porciani \& Madau}
\righthead{GRB Number Counts and Lensing Statistics}
\submitted{ApJ, in press}

\begin{document}
\onecolumn
\title{On the Association of Gamma--ray Bursts with Massive Stars:
Implications for Number Counts and Lensing Statistics}
\author{Cristiano Porciani\altaffilmark{1,2} and Piero Madau\altaffilmark{3,4}}

\altaffiltext{1}{Racah Institute of Physics, The Hebrew University of 
Jerusalem, Jerusalem, Israel.}
\altaffiltext{2}{Space Telescope Science Institute, 3700 San Martin
Drive, Baltimore, MD 21218.}
\altaffiltext{3}{Institute of Astronomy, Madingley Road, 
Cambridge CB3 0HA, UK.}
\altaffiltext{4}{Department of Astronomy and Astrophysics, University of 
California, Santa Cruz, CA 95064.}

\begin{abstract}
Recent evidence appears to link gamma--ray bursts (GRBs) to star--forming 
regions in galaxies at cosmological distances. If short--lived massive stars
are the progenitors of GRBs, the rate of events per unit cosmological 
volume should be an unbiased tracer (i.e. unaffected by dust obscuration and 
surface brightness limits) of the cosmic history of star formation. 
Here we use realistic estimates for the evolution of the stellar birthrate in 
galaxies to model the number counts, redshift distribution, and time--delay 
factors of GRBs. We present luminosity function fits to the BATSE $\log N-\log P$ 
relation for different redshift distributions of the bursts.
Our results imply about $1-2$ GRBs every one million Type II
supernovae, and a characteristic `isotropic--equivalent' burst luminosity
in the range $3-20\times 10^{51}\,$ergs s$^{-1}$ (for $H_0=65\,\kmsmpc$). 
We compute the rate of multiple imaging of background GRBs due to foreground  
mass condensations in a $\Lambda$--dominated cold dark matter cosmology, 
assuming that dark halos approximate singular isothermal spheres on galaxy 
scales and Navarro--Frenk--White profiles on group/cluster scales, and are 
distributed in mass according to the Press--Schechter model.
We show that the expected sensivity increase of 
{\it Swift} relative to BATSE could result in a few 
strongly lensed individual bursts detected down to a photon flux of
$0.1\,\punits$ in a 3--year survey. Because of the partial sky coverage,
however, it is unlikely that the {\it Swift} satellite will observe
recurrent  events (lensed pairs).

\noindent 
\end{abstract}
\keywords{cosmology: gravitational lensing -- theory -- gamma-rays: bursts}


\section{Introduction}
The Burst And Transient Source Experiment (BATSE) on the
{\it Compton Gamma Ray Observatory} ({\it CGRO}) has detected thousands
of gamma--ray bursts (GRBs) since 1991 (Paciesas \etal 1999). 
The distribution of BATSE bursts on the sky is isotropic, while the
intensity distribution shows a clear deficiency of
faint events relative to a uniform population of sources in Euclidean 
space (Meegan \etal 1992). Both these results provided the first clear
indication for a cosmological origin of GRBs. The discovery
of X--ray (Costa \etal 1997) and optical (van Paradijis \etal 1997)
afterglows has permitted to firmly establish the cosmological nature
of these events. 

From a theoretical perspective, the physical origin of GRBs is still uncertain.
Few known phenomena can release a suitable amount of energy to trigger such a 
powerful event, and  most models 
associate GRBs either with merging neutron stars or with the death of massive 
stars. Recent observations have provided some evidence that GRBs 
are related to star forming regions (Paczy{\'n}ski 1998; Fruchter 
\etal 1999),  and perhaps to some type of supernova explosion (Galama \etal 
1998). If the violent death of massive stars (whose lifetimes are much shorter
than the expansion timescale at the redshifts of interest) is somehow at the 
origin of the GRB phenomenon,  then the rate of events per unit cosmological 
volume should be an unbiased tracer -- unaffected by dust obscuration and 
surface brightness limits -- of the global star formation history of the 
universe. It has been pointed out by many authors (Lamb \& Reichart 2000; Blain
\& Natarajan 2000; Totani 1997, 1999; Krumholz, Thorsett, \& Harrison 1999; 
Lloyd \& Petrosian 1999; Wijers \etal 1998; Sahu \etal 1997) that 
standard statistical analyses of GRBs and their afterglows could then be used 
to derive additional constraints on the evolution of the stellar birthrate
(SFR), and to gain further insight on the nature of these events.

In this paper we study the expected cosmological distribution of GRBs
in the massive star progenitor scenario, identify some uncertanties in the 
data and in their interpretation, and discuss future 
observations for addressing them.
We show that the brightness distribution of the BATSE bursts can be well 
reproduced by assuming a proportionality between the GRB rate density 
and observationally--based SFR estimates, once the standard candle hypothesis
is relaxed (cf. Totani 1999). By itself, the bursts number--flux
relation cannot discriminate between different plausible star formation 
histories (see Krumholz \etal 1998) since, given a SFR and assuming a functional 
form for the intrinsic luminosity function of GRBs, the values of free 
parameters can always be optimized to reproduce the observed number counts.
On the other hand, quantities which reflect the redshift distribution of
the bursts (like, e.g., the ratio between the average durations of 
bright and faint events) do depend on the underlying SFR and could be
used as discriminants. With this problem in mind we also reassess the detectability of multiply--imaged 
GRBs due to the strong lensing effect of foreground mass concentrations 
(Paczy\'{n}ski 1986; Mao 1992).
Events associated with galaxy (cluster) lenses will produce images with typical
angular separations of a few ($\sim 20$) arcseconds and time delays of 
the order of weeks or months (years). Multiply--imaged bursts cannot be 
spatially resolved by present--day gamma--ray detectors and will appear as 
`mirror' or recurrent events -- same location on the sky, identical spectra 
and light curves -- at different times and with different intensities. 
While a number of strongly lensed individual bursts could be detected by 
{\it Swift}, the restricted sky coverage makes the probability of observing 
a lensed pair rather small.

\section{log $N$--log $P$ distribution}

The photon flux (in units of $\punits$) observed at Earth in the 
energy band $E_{\rm min}<E<E_{\rm max}$ and emitted by an isotropically 
radiating source at redshift $z$ is
\be
P=\f{\displaystyle (1+z) \int_{(1+z)E_{\rm min}}^{(1+z)E_{\rm max}}S(E) \,dE}
{4 \pi d^2_{\rm L}(z)}\;,
\ee
where $S(E)$ is the differential rest--frame photon luminosity of the 
source (in units of s$^{-1}\,$ keV$^{-1}$), and $d_L(z)$ is the standard 
luminosity distance for a  Friedmann--Robertson--Walker
(FRW) metric. It is customary to define an `isotropic equivalent' burst 
luminosity in the energy band 30--2000 keV as $L=\int_{30 \,{\rm keV}}
^{2000 \,{\rm keV}} E \,S(E) \,dE$. If we denote with $\psi(L)$ the GRB 
luminosity function (normalized to unity), then the {\it observed} rate of 
bursts with {\it observed} peak fluxes in the interval $(P_1,P_2)$ is 
\be
{dN\over dt}(P_1 \leq P< P_2)=\int_0^\infty
 dz \,\f{dV(z)}{dz} \f{R_{\rm GRB}(z)}{1+z} 
\int_{L(P_1,z)}^{L(P_2,z)} dL'\, \psi(L') \epsilon(P)\;,
\label{counts1}
\ee
where $dV/dz$ is the comoving volume element, $R_{\rm GRB}(z)$ is the 
comoving GRB rate density, $\epsilon(P)$ is the detector efficiency as a function 
of photon flux, and the factor $(1+z)^{-1}$ accounts for cosmological time 
dilation. If the geometry of the universe is FRW on large scales, then 
\be
{dV\over dz}= {c\over H_0}\f{\Delta\Omega_s d_L^2(z)}
{(1+z)^2 [\Om_M (1+z)^3+\Om_K(1+z)^2+\Om_\Lambda]^{1/2}},
\ee
where $\Delta\Omega_s$ is the solid angle covered on the sky by the survey, and
$\Omega_K=1-\Om_M-\Om_{\Lambda}$ is the curvature contribution to
the present density parameter.
Unless otherwise stated, we shall assume in the following a vacuum--dominated
cosmology with density parameters $\Omega_M=0.3$ and $\Om_\Lambda=0.7$,
and a Hubble constant $H_0=65 \,h_{65}\kmsmpc$.

\section{Star formation history}

Our starting hypothesis is that the rate of GRBs traces the global star 
formation history 
of the universe, $R_{\rm GRB}(z)\propto R_{\rm SF}(z) \propto R_{\rm SN}(z)$,
with $R_{\rm SF}$ and $R_{\rm SN}$ the comoving rate densities of star 
formation and core--collapse (Type II) supernovae, respectively. The constant 
of proportionality, $k\equiv R_{\rm SN}/R_{\rm GRB}$, is a free--parameter of 
the model. Popular scenarios for GRBs include merging neutron stars
(Paczy{\'n}ski 1986) or the formation of black holes in supernova--like events 
(`collapsars', MacFadyen \& Woosley 1999). The key idea here is to assume that 
GRBs are produced by stellar systems which evolve rapidly -- by cosmological 
standards -- from their formation to the explosion epoch. This would not be
true at high redshift in the case of coalescing neutron stars, which have 
a median merger time of 100 Myr according to the recent population 
synthesis study of Bloom, Sigurdsson, \& Pols (1999). 

\begin{figure}
\epsscale{0.4}\plotone{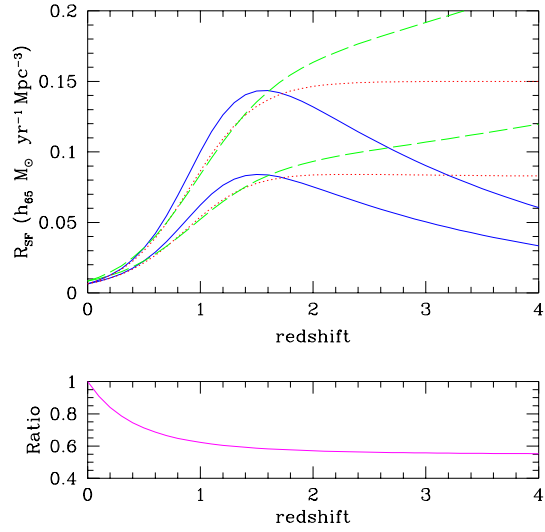}
\caption{Cosmic star formation history. The continuous, dotted, and dashed 
curves show the rate of star formation per unit comoving volume
as a function of redshift, for our models SF1, SF2, and SF3, respectively.
The top set of curves refers to an EdS universe,
while the lower lines are for a $\Lambda$--dominated cosmology.
The ratio between the two sets is shown in the small panel at the bottom.
}
\label{SFR}
\end{figure}

A number of workers have modelled the expected evolution of the cosmic SFR 
with 
redshift. Most have followed a similar route to that employed by Madau \etal
(1996), who based their estimates on the observed 
(rest--frame) UV luminosity density of the galaxy population as a whole. Using 
various diagnostics, the cosmic SFR can now be traced to $z\approx 4$, 
although some details remain controversial. We use here three different 
parameterizations (shown in Figure \ref{SFR}) of the global star formation 
rate 
per unit comoving volume in an Einstein--de Sitter universe (EdS). The first 
(hereafter SF1) is taken from Madau \& Pozzetti (2000): 
\be
R_{\rm SF1}(z)=0.3 \, h_{65} \,\f{\displaystyle{\exp(3.4z)}}{\displaystyle
{\exp(3.8z)+45}}\sfrd\;.
\label{SF1}
\ee
This star formation history matches most measured UV--continuum and H$\alpha$ 
luminosity 
densities, and includes an upward correction for dust reddening of 
$A_{1500}=1.2$ mag. The SFR increases rapidly between $z=0$ and $z=1$, peaks 
between $z=1$ and 2, and gently declines at higher redshifts. 
Because of the uncertainties associated with the incompleteness of
the data sets and the amount of dust extinction at early epochs, we consider a 
second scenario in which the SFR remains instead roughly constant at $z\gta 
2$ (Steidel \etal 1999),
\be
R_{\rm SF2}(z)=0.15\, h_{65} \,
\f{\displaystyle{\exp(3.4z)}}{\displaystyle{\exp(3.4z)+22}}
\sfrd
\label{SF2}
\ee
(SF2). Some recent studies have suggested that the evolution of the SFR up to 
$z\approx 1$ may have been overestimated (e.g. Cowie, Songaila, \& Barger 
1999), while the rates at high--$z$ may have been severely underestimated 
due to large amounts of dust extinction (e.g. Blain \etal 1999). We then 
consider a third SFR,
\be
R_{\rm SF3}(z)=0.2\, h_{65} \,
\f{\displaystyle{\exp(3.05 z-0.4)}}{\displaystyle{\exp(2.93z)+15}}
\sfrd 
\label{SF3}
\ee
(SF3), with more star formation at early epochs.
In every case we adopt a Salpeter initial mass function (IMF) -- assumed 
to remain constant with time -- with a lower cutoff around $0.5\,M_\odot$ 
(Madau \& Pozzetti 2000), consistent with observations of M subdwarf disk 
stars (Gould, Bahcall, \& Flynn 1996). A constant multiplicative factor of
1.67 will convert the SFR to a Salpeter IMF with a cutoff of $0.1\,M_\odot$. 
To include the effect of a $\Lambda$--dominated cosmology we have computed the 
difference in luminosity density between an EdS and a $\Lambda$ universe,
and applied this correction to the SFR above (see Appendix A for details). 
Assuming that all stars with masses $M>8\,{\rm M}_\odot$ explode 
as core--collapse supernovae (SNe), the SN II rate density $R_{\rm SN}
(z)$ can then be estimated by
multiplying the selected SFR by the coefficient 
\be
{\int_8^{125} dm \,\phi(m)\over \int_{0}^{125} 
dm \, m\, \phi(m)}= 0.0122 \,{\rm M}_\odot^{-1}\;,
\ee
where $\phi(m)$ is the IMF and $m$ the stellar mass in solar units. The 
resulting 
rates agree within the errors with the locally observed value of $R_{\rm 
SN}=(1.1 \pm 0.4) \times 10^{-4}\,h_{65}^3 \,{\rm yr}^{-1}\, {\rm Mpc}^{-3}$ (e.g. Madau, 
della Valle, \& Panagia 1998 and references therein).

\section{Luminosity function}

The observed fluxes from GRBs with secure redshifts
rule out the classical standard candle hypothesis (see Table 1 of Lamb \& 
Reichart 2000 and references therein): the inferred `isotropic--equivalent' 
photon luminosities at peak vary by about a  factor of 50, with a mean value 
of $3.8\, h_{65}^{-2} \times 10^{58} \,{\rm s}^{-1}$.
The data are too sparse, however, for an empirical determination of
the burst luminosity function, $\psi(L)$. To model the number counts 
we then simply assume that the burst luminosity distribution does not evolve 
with redshift and adopt a simple functional form for $\psi(L)$, 
\be
\psi(L)=C \, \left(\f{L}{L_0}\right)^{\gamma} \exp\left(-\f{L_0}{L}\right),
\label{LF}
\ee
where $L$ denotes the peak luminosity in the 30--2000 keV energy range 
(rest--frame), $\gamma$ is the asymptotic slope at the bright end, $L_0$
marks a 
characteristic cutoff scale, and the constant $C=[L_0\Gamma(-\gamma-1)]^{-1}$ 
(for $\gamma<-1$) ensures a proper normalization $\int_0^\infty \psi(L) dL=1$.


\section{Photon spectrum}

To describe the typical burst spectrum we adopt the functional form empirically
proposed by Band \etal (1993):
\be
S(E)=A\times  
\left\{ \begin{array}{ll}
\displaystyle{
\left(\f{E}{100\,{\rm keV}}\right)^\alpha\exp\left[\f{E
(\beta-\alpha)}{E_b}\right]}
& \textrm{if $E<E_b$}\;;\\
\displaystyle{
\left(\f{E_b}{100\,{\rm keV}}\right)^{\alpha-\beta}
\exp{(\beta-\alpha)} 
\left(\f{E}{100 \,{\rm keV}}\right)^\beta}
& \textrm{if $E\geq E_b$}\;.
\end{array} \right.
\ee
For simplicity, the low and high energy spectral indices, $\alpha$ and 
$\beta$, have been assigned the values of $-1$ and $-2.25$, respectively, for
all bursts. These are the mean values recently measured by Preece \etal 
(2000) for a large collection of bright BATSE events. The assumed 
characteristic 
energy of the spectral break is $E_b=511$ keV. 
Note that the method introduced by Fenimore \& Bloom (1995) to account for
the spectral diversity of bursts by averaging the number counts over
the spectral catalog by Band \etal (1993) cannot be rigorously applied
when the standard candle hypothesis is relaxed.
We have checked, a posteriori, the stability of our results 
with respect to small variations of the spectral parameters. 
This issue is briefly discussed in the next session.

\section{Comparison with the data}
\begin{deluxetable}{c c c c c c c c c c c}
\footnotesize
\tablenum{1}
\tablecaption{Best--fitting parameters for the $\log N - \log P $ relation.}
\tablewidth{0 pt}
\tablehead{
\colhead{Model} & $\gamma$ & $L_0 (10^{51} h^{-2}_{65} {\rm erg \,s}^{-1})$ 
& $k_{\rm B} (10^5 h^{-2}_{65})$ & $\chi^2_{\rm B}$ & $k_{\rm P}
(10^5 h^{-2}_{65})$ & $\chi^2_{\rm P}$}
\startdata
SF1  &  $-2.5 \pm 0.2$ & $3.2 ^{+1.0}_{-0.7}$ & $ 4.4 \pm 0.2$ & $18.5 $ & 
$ 1.82 ^{+0.17}_{-0.14}$ & $ 6.26$ \nl
SF2  &  $-2.9^{+0.4}_{-0.6}$ & $7^{+4}_{-2} $ &  $6.0^{+0.4}_{-0.3}$ & 
$ 20.9 $ & $ 2.17^{+0.20}_{-0.17}$ & $5.69$ \nl
SF3  &  $-3.7^{+0.8}_{-2.2}$ & $17^{+19}_{-8}$ & $8.2^{+0.4}_{-0.5} $ & 
$23.2 $ & $ 2.82 ^{+0.26}_{-0.22} $ & $ 5.70$ \nl 
\enddata
\label{fit}
\end{deluxetable}
To calibrate and test our models against the observed number counts we have
used the off--line BATSE sample of Kommers \etal (2000), which includes 1998  
archival BATSE (``triggered'' plus ``non--triggered'') bursts in the energy 
band 50--300 
keV. The efficiency of this off--line search is well--described
by the function $\epsilon(P)=0.5[(1+{\rm erf}(-4.801+29.868 P)]$ 
(Kommers \etal 2000).

\begin{figure}
\epsscale{0.4} \plotone{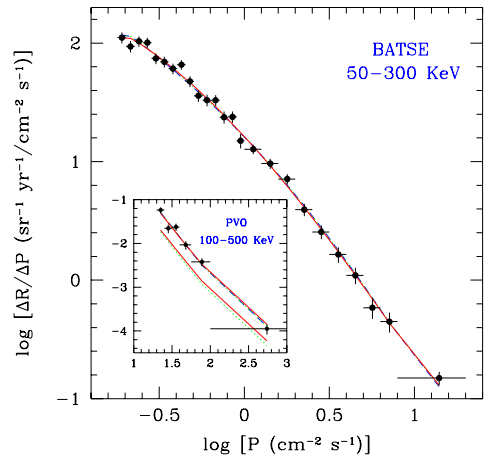}
\caption{Differential GRB number counts versus peak photon flux.
The points and vertical error bars show the observed rates from Kommers \etal 
(2000) and their Poisson uncertainties (horizontal error bars denote bin
sizes). The best--fit models obtained by assuming that the burst rate is
proportional to SF1, SF2 and SF3 are shown with (overlapping) solid, dotted,
and dashed lines, respectively. The comparison of the model results with 
{\it PVO} rates is depicted in the inset. In this case two sets of curves 
are plotted: the upper one represents the best fit to the {\it PVO} data, 
the lower set the extrapolation of the rates shown in the main panel.}
\label{cnts}
\end{figure}

We have optimized the value of our three free parameters, $k, \gamma$ and 
$L_0$, by $\chi^2$ minimization over 25 peak flux intervals 
(see Table 2 in Kommers \etal 2000).
In Figure \ref{cnts} we show the best--fitting results for the three different
star formation histories considered. The observed number counts have been 
converted into rates per unit time per unit solid angle
by estimating the effective live time of the searches and their
field of view following Kommers \etal (2000). 
Assuming a normal distribution for the errors, we can relate
confidence levels to value intervals for the free parameters.
Table \ref{fit} gives the chi--square of the best--fitting models,
$\chi^2_{\rm B}$, 
the ranges corresponding to the $68 \%$ confidence level
for the parameters that determine the luminosity function ($\gamma$
and $L_0$), and the expected number of core--collapse supernovae 
per BATSE burst, $k_{\rm B}$. 

The overall quality of the best--fit decreases when the star formation rate
at high redshift is increased, i.e. as the models start predicting too many 
bursts to be consistent with the faintest off--line BATSE counts: the minimum 
$\chi^2$ per degree of freedom 
is $18.5/22=0.84$ for $R_{\rm SF1}$, $20.9/22=0.95$ for $R_{\rm SR2}$, and 
$23.2/22=1.05$ for $R_{\rm SF3}$. 
\begin{deluxetable}{c c c c c c}
\footnotesize
\tablenum{2}
\tablecaption{Moments of the best--fitting luminosity functions  
(all in units of $10^{51} h^{-2}_{65} {\rm erg \,s}^{-1}$).
\label{ddelay}}
\tablewidth{0 pt}
\tablehead{
\colhead{Model} & $L_5$ & $L_{95}$ 
& $\langle L \rangle $ & $L_{50}$ & $L_+$
}
\startdata
SF1  &  $0.82$ & $18.19$ & $ 6.40$ & $2.71 $ & $ 1.28 $  
\nl
SF2  &  $1.57$ & $22.78$ & $ 8.00$ & $4.56 $ & $ 2.48 $
\nl
SF3  &  $2.82$ & $24.71$ & $ 9.71$ & $6.95 $ & $ 4.46 $
\nl 
\enddata
\label{lum}
\end{deluxetable}
Strong covariance of $\gamma$ and $L_0$ is observed
in the region of parameter space surrounding the best--fitting values.
When the slope of the high--luminosity tail of the $\psi(L)$ is increased,
one has to correspondingly raise the value of the cutoff luminosity to 
prevent a strong $\chi^2$ increment: this means that our models need the 
presence of relatively high--luminosity events to reproduce the data.
Luminosity intervals corresponding to $90 \%$ of the bursts (obtained excluding
the $5 \%$ tails on both sides) are given in Table \ref{lum}.
The average ($\langle L\rangle$), median ($L_{50}$), and mode ($L_+$) 
of the distributions are also given
\footnote{Note that $\langle L^n \rangle=L_0^n \,\Gamma(-\gamma-n-1)/
\Gamma(-\gamma-1)$ if $\gamma<-(n+1)$, and diverges otherwise.}.
The derived rate of GRBs at the present epoch ranges from $(0.181\pm 0.008)\, h_{65}^
{-1}\,{\rm yr}^{-1}\,{\rm Gpc}^{-3}$ (SF1) down to $(0.133^{+0.006}_
{-0.009})\, 
h_{65}^{-1}  \,{\rm yr}^{-1}\,{\rm Gpc}^{-3}$ (SF2) and $(0.125^{+0.008}
_{-0.006})\, 
h_{65}^{-1}  \,{\rm yr}^{-1}\,{\rm Gpc}^{-3}$ (SF3).
The best--fit values for $\gamma$ and $L_0$ are found to be rather insensitive 
to $20\%$ variations in the spectral parameters $\alpha$ and $E_0$.
The cutoff scale of the luminosity function, however, 
depends more sensitively on the assumed value of the high--energy spectral
index $\beta$ (this is especially true for $\beta>-2.25$).
For example, for SF1 and $\beta=-2.0$, the best--fitting parameters become
$L_0=3.7 ^{+1.0}_{-0.8}$ and $\gamma=-2.5 \pm 0.2$.

It is of interest to compare the properties of the luminosity functions
that provide the best--fit for each star formation rate. 
As expected, to balance the effect of cosmic expansion,
the typical burst luminosity increases in models with larger amounts of 
star formation at early epochs (see Table \ref{lum}).
Moreover, changing from SF1 to SF3, the luminosity function broadens,
becoming less and less peaked around $L_+$, while
the slope of the luminosity function
in the range $L_+ \lta L \lta L_{95}$, $\gamma_{\rm eff}$,
remains the same in all the models, $\gamma_{\rm eff}\sim 2.3$. 
An increase in the amount
of star--formation at high redshifts requires a steeper
high--luminosity tail of $\psi(L)$.
Regrettably,
since the number of GRBs with known redshift is very small,
it is not yet possible to use observational data to discriminate 
among different luminosity functions (see below, however, for 
a comparison between the redshift distributions of bursts 
predicted by our models as a function of measured peak flux, and the 
available data).

\begin{deluxetable}{c c c c c c c c c c c}
\footnotesize
\tablenum{3}
\tablecaption{Average redshifts and time--delay factors.  \label{delay}}
\tablewidth{0 pt}
\tablehead{
\colhead{Model} & $\langle z \rangle_{\rm b}$ &  
$\langle z \rangle_{\rm i}$ & 
$\langle z \rangle_{\rm f}$ & 
$\langle z \rangle_{\rm t}$
& $\langle z\rangle _{\rm Nb}$ &
$\langle z\rangle _{\rm Nd}$ &
$\displaystyle{\f{
\langle 1+z\rangle _{\rm Nd}}{\langle 1+z\rangle _{\rm Nb}}}$ &
$\langle z\rangle _{\rm Nb'}$ &
$\langle z\rangle _{\rm Nd'}$ &
$\displaystyle{\f{
\langle 1+z\rangle _{\rm Nd'}}{\langle 1+z\rangle _{\rm Nb'}}}$
}
\startdata
SF1  &  $0.94$ & $ 1.45$ & $ 2.29 $ & $ 1.67$ &
 $0.99 $ & $1.59$ & $1.30$ & $0.90$ & $1.58$ & $1.36$\nl
SF2  &  $0.92$ & $ 1.72$ & $ 3.10 $ & $ 2.08$ &
 $ 1.00$ & $1.94$ & $1.47$ & $0.85$ & $1.92$ & $1.57$\nl
SF3  &  $0.85$ & $ 1.89$ & $ 3.71 $ & $ 2.37$ & 
 $ 0.96$ & $2.19$ & $1.63$ & $0.77$ & $2.16$ & $1.79$\nl 
\enddata
\label{shift}
\end{deluxetable}

To test our models against observations of very bright and rare 
bursts we have also used the number counts accumulated by the 
{\it Pioneer Venus Orbiter} ({\it PVO}) at $20<P<1000\,\punits$ 
in the 100--500 keV band (see Table 2 in Fenimore \& Bloom 1995).
Since no threshold effects are expected in the {\it PVO} detection of such bright 
events, we set $\epsilon(P)=1$ in equation (\ref{counts1}) for 
the counts. By combining {\it PVO} and BATSE data we should be able to test our 
models over about 3.5 orders of magnitudes in peak flux. We find that, while
our best--fitting models for the BATSE counts
have the right shape to accurately describe the {\it PVO} rates (roughly a $3/2$
power--law), the predicted counts need to be multiplied by a factor 
of $\sim 2.5-3$ to have the right normalization (see Figure \ref{cnts}). To 
better
quantify this discrepancy, we have minimized the chi--square function
using only {\it PVO} data (divided in 6 bins as in Fenimore \& Bloom
1995), and allowing just the parameter $k$ to vary, while assigning
to $\gamma$ and $L_0$ the values given in Table \ref{fit}. The resulting 
minimum chi--square, $\chi^2_{\rm P}$, and the corresponding normalization 
constant of the GRB rate, $k_{\rm P}$, are shown in Table \ref{fit}.
It is possible that the {\it PVO} and off--line BATSE catalogues,
using different selection criteria, may not form a homogeneous burst sample.
\footnote{Note that our models, when normalized to fit the {\it PVO} data, 
automatically account for the on--line BATSE counts given in Table 2 of 
Fenimore \& Bloom (1995); these have been carefully selected to be 
consistent with the {\it PVO} data set.}~ 
The {\it PVO} catalogue does not report the trigger time--scale for burst 
detection, and each light curve has
been analyzed a posteriori to select only events above 
the detection threshold on timescales of either 0.25 s or 1 s.
Kommers \etal (2000) have included only events detected on a timescale of 1.024
s: short--duration bursts may then be under--represented in the off--line BATSE 
sample. Alternatively, the discrepancy could be explained with the existence 
of a local (bright) population of GRBs.
In the following we will only include long--duration bursts in our analysis 
and use models calibrated against archival BATSE data.

\begin{figure}
\epsscale{0.5} \plotone{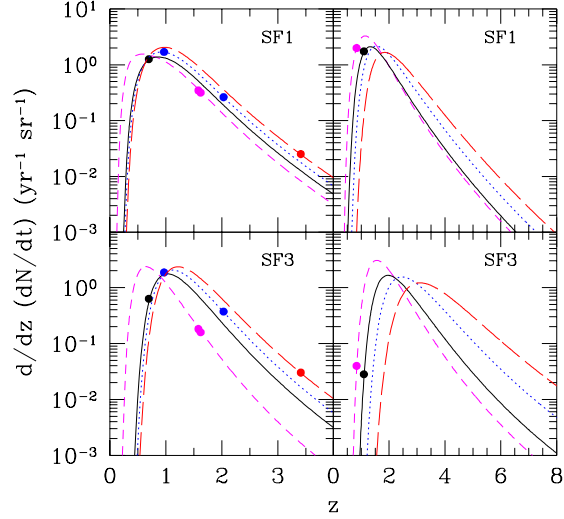}
\caption{Redshift distributions of the bursts detected in the
off--line search by Kommers \etal (2000). In the left panels the short--dashed, 
solid, dotted, and long--dashed curves refer, respectively, to the
peak flux ($\punits$) intervals $(7.943,20.00)$, $(3.162,3.981)$, 
$(2.511,3.162)$, and $(1.995,2.511)$. The corresponding intervals in the right 
panels are $(1.000,1.259)$, $(0.569,0.639)$, $(0.320,0.359)$, and 
$(0.180,0.202)$. The curves in the upper (lower) panels have been derived
assuming a burst rate proportional to SF1 (SF3). GRBs with known redshifts
have been plotted as filled points on the curve corresponding to 
their observed brightness.}
\label{zgrb}
\end{figure}

In Figure \ref{zgrb}, the expected redshift distribution of bursts, 
$d/dz (dN/dt)$ (with $dN/dt$ defined in 
eq. \ref{counts1}), is plotted as a function of redshift for a number 
of selected luminosity intervals (the efficiency of BATSE off--line search is 
assumed), and two different star formation histories: SF1 (upper panels) and
SF3 (lower panels).
Bright events ($P \geq 1 \punits$) are depicted in the left panels, faint 
bursts on the right.
The peak flux intervals used by Kommers \etal (2000) (approximately evenly 
spaced in $\log P$) 
which contain at least one GRB with known redshift are considered.
A similar analysis for the faint bursts ($P < 1 \punits$) is performed in 
the right panels. In this case, we also plot the redshift distribution
of the sources corresponding to two luminosity intervals in which 
no afterglow redshifts have been determined.
The redshifts of bursts with known optical 
counterparts (see Table 1 of Lamb \& Reichart 2000), including GRB000301C
(Smith, Hurley, \& Kline 2000; Castro \etal 2000)
are also shown as filled points on the curve corresponding to 
their observed brightness.
We do not include GRB980425 in this analysis, since its association
with SN 1998 bw (a Type Ic at $z=0.0085$) is uncertain (Pian \etal 2000).
It might well be the case that GRB980425 is representative of a special
class of GRBs (e.g. Bloom \etal 1998).
 
Note that, barring selection effects, model SF1 can 
reasonably account for all but the highest observed redshift ($z=3.418$ for 
GRB971214), which has a low a priori probability in all three models
(cf. Schmidt 1999). 
The detection of relatively faint bursts at $z\sim 1$ (like GRB970508 and 
GRB980613) appears improbable in model SF3. Table \ref{shift} 
lists the expected average 
redshift of GRBs observed in three selected flux ranges: a bright sample 
($7.94<P<20\,\punits$, subscript $b$), an intermediate sample ($1.0<P<2.26\,
\punits$, subscript $i$), and a faint sample ($0.18<P<0.20\,\punits$, 
subscript $f$). The entire interval $0.18<P<20\,\punits$ is also considered 
(subscript $t$). The redshift distribution of GRBs has another direct 
observational consequence: because of cosmic expansion,
faint bursts will have longer durations than bright ones, on the average.
This time dilation effect is proportional to $1+z$.
Even though observational results are still controversial, a 
cosmological time dilation 
factor of about 2 between dim and bright bursts is widely accepted for the 
long duration events (Norris \etal 1994, 1995; Norris 1996).
The average redshifts of the bursts lying in the peak flux intervals considered
by Norris \etal (1995) are also shown in Table \ref{shift}. 
The quantities $\langle z \rangle_{\rm Nb}$ and 
$\langle z \rangle_{\rm Nd}$ refer, respectively, to their bright 
($4.21\leq P \leq 58.28\, \punits$)
and combined dim + dimmest ($0.33\leq P \leq 2.82\, \punits$) samples 
(see Table 1 in Horack, Mallozzi \& Koshut 1996).
Since, for both classes of GRBs, the peak flux distribution of the dataset 
used in the analysis of
Norris \etal (1995) is strongly peaked around the mean value, we repeated
our calculations considering smaller $P$ intervals.
The quantities $\langle z \rangle_{\rm Nb'}$ and 
$\langle z \rangle_{\rm Nd'}$ are, in fact, performed over the ranges
of peak flux corresponding to $\langle P \rangle \pm 2 \sigma_{\langle P
\rangle}$ (i.e. $12.34 \leq P \leq 20.06\,\punits$ for the bright sample, and
$0.72\leq P \leq 0.84\,\punits$ for the faint one), with $\sigma_{\langle P
\rangle}$ the standard deviation of the mean.  The time dilation 
estimates of $\sim 2.25$ (Norris \etal 1995) and $\sim 1.75$ (Norris 1995) 
appear to favor scenarios in which the SFR does not decrease at 
high--redshift. On the other hand,
SF1 is slightly preferred to the other cosmic star formation histories
by the $\,\log N-\log P$ analysis of the BATSE data (see the $\chi^2_{\rm B}$
values in Table \ref{fit}).
A recent search for non--triggered GRBs in the BATSE records (Stern \etal 2000),
however, has 
detected many more faint bursts than the analysis by Kommers \etal (2000).
The ratio between the estimated number counts can be as high as a 
factor of $\sim 2$ at $P=0.2 \punits $. If confirmed, a large number of 
faint counts would probably favor scenarios in which the star--formation 
rate does not substantially decrease at $z\gta 3-4$.   

\section{Gravitational lensing of GRBs}

In the observable (clumpy) universe, gravitational lensing will magnify and 
demagnify high redshift GRBs relative to the predictions of ideal (homogeneous)
reference cosmological models. A burst that goes off within the Einstein ring 
of a foreground mass concentration may generate multiple images at different 
positions on the sky. The magnitude and frequency of the effect depend on 
the redshift distribution of the sources,
the abundance and the clustering properties of virialized clumps, the mass 
distribution within individual 
lenses, and the underlying world model. Events associated with galaxy 
(cluster) lenses will produce images with typical angular separations of a 
few ($\sim 20$) arcseconds, smaller than the presently achievable 
$\gamma$--ray instrumental resolution. On the other hand, GRBs are transient 
phenomena with durations ranging from a fraction of a second to several 
hundreds seconds, while the typical time delay between multiple images
is of the order of weeks in the case of galaxy lensing, and years for lensing
by a foreground cluster. Mirror images of the same burst will then 
appear as separate events with overlapping positional error boxes,
identical time histories, and intensities that differ only by a scale factor.
In principle, the detection of two or more images satisfying these three 
conditions should pinpoint a good candidate for a lensed GRB.
However, temporal variation of the background signal and the 
presence of noise in observed light curves can make this task estremely
difficult (Wambsganss 1993), and special statistical methods devised to 
minimize the effect of the noise must be adopted for light curve comparison 
(e.g. Nowak \& Grossman 1994).

In this section we estimate the number of multiply--imaged GRBs expected
as a function of the limiting
sensitivity of the survey, and for the different star formation histories  
discussed in \S\,3. We improve upon previous calculations of GRB lensing 
by using more realistic models of the burst redshift and brightness
distributions, and of the foreground mass concentrations.    
Following our previous study on
high--$z$ supernovae (Porciani \& Madau 2000),
we assume that lensing events are caused by intervening dark matter halos 
which approximate singular isothermal spheres on galaxy scales 
and Navarro--Frenk--White (Navarro, Frenk, \& White 1997; hereafter NFW)
profiles on group/cluster scales, and are distributed 
in mass according to the Press--Schechter (Press \& Schechter 1974; hereafter 
PS) theory.
This model for the lens population provides a good fit to the data
on QSO image separations lensing, and may originate in a scheme 
which includes the dissipation and cooling of the baryonic protogalactic 
component and the radial re--distribution of the collisionless dark matter as
a consequence of baryonic infall (e.g. Keeton 1998).
The strong lensing optical depth for a light--beam emitted by a point source
at redshift $z_s$ is (Turner, Ostriker, \& Gott 1984)
\be
\tau(z_s)=\int_0^{z_s} dz (1+z)^3
\f{dl}{dz}
\int_0^\infty dM \,\Sigma(M,z,z_s) n(M,z)\;,
\label{tautog}
\ee
where $dl/dz=c H_0^{-1} (1+z)^{-1} [\Om_M(1+z)^3+\Om_K(1+z)^2+\Om_{\Lambda}]^
{-1/2}$ is the cosmological line element, $\Sigma(M,z,z_s)$ 
is the lensing cross 
section as measured on the lens--plane, and $n(M,z)$ the
comoving differential distribution of halos with mass $M$ at redshift $z$.
Equation (\ref{tautog})
assumes that each bundle of light rays encounters only one lens, the
lens population is randomly distributed, and the resulting $\tau\ll 1$.
The mass distribution in a single lens and the geometry of the
source--lens--observer system completely determine {\bf$\Sigma(M,z,z_s)$}.
For a singular isothermal sphere (SIS) with one--dimensional velocity 
dispersion $\s_v$, the strong lensing cross section is 
\be
\Sigma_{\rm SIS}(\s_v,z_s,z_l) = 16 \pi^3 \left(
\f{\s_v}{c}\right)^4 \left(\f{D_lD_{ls}}{D_{s}} \right)^2 \;,
\label{crsecsis}
\ee 
where $D_l,D_s,$ and $D_{ls}$ are the angular diameter distances 
between the observer--lens, the observer--source, and the lens--source systems.
According to the PS theory, the differential comoving number 
density of dark halos with mass $M$ at redshift $z$ is given by 
\be
n(M,z)=\f{1}{\sqrt{2 \pi}}\f{\rho_0}{M}\f{\de_c(z)}{\s_M^3}\exp\left[-
\f{\de_c^2(z)}{2 \s_M^2} \right] \left| \f{d\s_M^2}{dM}\right |\;,
\label{ps}
\ee
where $\rho_0$ is the present mean density of the universe. The halo abundance
is then fully determined  by
the redshift--dependent critical overdensity $\de_c$ (e.g. Eke, Cole, \& 
Frenk 1996) and by the linearly extrapolated (to $z=0$)
variance of the mass--density field smoothed on the scale $M$, $\s_M^2$.
The latter is computed assuming
a scale--invariant power--spectrum of primordial density fluctuations
with spectral index $n_{\rm p}=0.96$ 
and the transfer function for CDM given in Bardeen \etal (1986).
The amplitude of density perturbations is fixed by requiring 
the (present--day, linearly extrapolated) {\it rms} 
mass fluctuation in a $8 \; h^{-1}$ Mpc sphere to be $\sigma_8=0.87$. 

Assuming that every halo virializes to form a (truncated) singular isothermal
sphere of velocity dispersion $\s_v$, mass conservation implies
\be
\s_v(M,z)=\f{1}{2} H_0 \left(\f{3 M}{4\pi\rho_0}\right)^{1/3}\,\Omega_M^{1/3}
\Delta^{1/6}\left[\f{\Omega_M}{\Omega(z)}\right]^{1/6}(1+z)^{1/2} \;,
\label{virial}
\ee
where $\Omega(z)=\Omega_M (1+z)^3/[\Omega_M (1+z)^3+\Omega_K (1+z)^2
+\Omega_\Lambda]$. Here $z$ denotes the virialization epoch of the halo,
and $\Delta(z)$ is the ratio between its actual mean density at
virialization and the corresponding critical density, $\rho_{\rm crit}(z)=
3 H^2(z)/8 \pi G$ (here $H(z)$ the Hubble parameter at redshift $z$, and $G$ 
the gravitational constant). Equation (\ref{virial}) 
relates the PS mass function to the SIS lens profile, therefore allowing the 
computation of the optical depth given in equation (\ref {tautog}).
For the NFW density profile -- shallower than isothermal near the halo center 
and steeper than isothermal in its outer regions -- the lens equation must 
be solved numerically. With respect to a halo SIS profile containing the same 
total mass, a NFW lens has a smaller cross section for multiple imaging, but 
generates a higher magnification.

\begin{figure}
\epsscale{0.4} \plotone{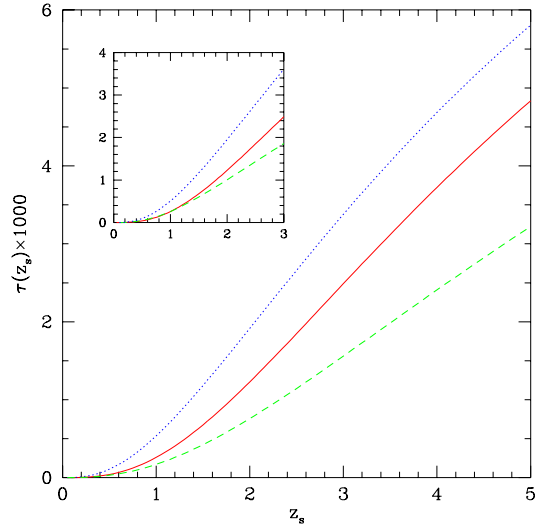}
\caption{Strong lensing optical depths for a point source at $z_s$ in three
different hierarchical cosmogonies. The mass distribution of the lenses
is described by the Press--Schechter theory; a lens having mass $M$ is 
modelled by a singular isothermal sphere for $M<3.5\times 10^{13} M_\odot$, and by a 
NFW profile otherwise. {\it Solid line:} $\Lambda$CDM.
{\it Dashed line:} OCDM. {\it Dotted line:} SCDM.
In the inset, the $\Lambda$CDM optical depth ({\it solid curve}) is compared
with the optical depth due to the known population of elliptical galaxies. 
This is obtained
by extrapolating the locally observed galaxy population to higher redshifts,
assuming a constant comoving number density of ellipticals
(e.g. Kochanek 1993; Maoz \& Rix 1993). The dashed and dotted lines 
correspond to the luminosity function 
of Marzke \etal (1998) and Ellis \etal (1996), respectively. 
See Porciani \& Madau (2000) for a detailed description of the model.
}
\label{tau}
\end{figure}

The resulting optical depths for strong lensing are plotted in Figure \ref{tau}
for our reference cosmology ($\Lambda$CDM) and two other popular
cold dark matter models: OCDM ($\Om_M=0.3$, $\Om_\Lambda=0$, $n_{\rm p}=1$, 
$\sigma_8=0.85$, and $H_0=70\,\kmsmpc$), and 
SCDM ($\Om_M=1$, $\Om_\Lambda=0$, $n_{\rm p}=1$, $\sigma_8=0.5$, 
and $H_0=50\,\kmsmpc$).
In all cases the amplitude of the power spectrum has been fixed in order to 
reproduce the observed
abundance of rich galaxy clusters in the local universe (e.g. Eke, Cole \&
Frenk 1996). In $\Lambda$CDM a convenient fit to the lensing
optical depth is
\be
\tau(z)={8.4\times 10^{-4}\over 3.1z^{-2.85}-0.39z^{-1.42}+z^{-1}+
9.5 \times 10^{-4} z^{1.5}},
\ee
to within 1\% in the range $0.6<z<7$.
Our analytical method is expected to be very accurate
for sources at $z_s\leq 3$, and to slightly underestimate the optical
depth for multiple lensing at higher redshift (Holz, Miller, \& Quashnock
1999). 

\begin{figure}
\epsscale{0.4} \plotone{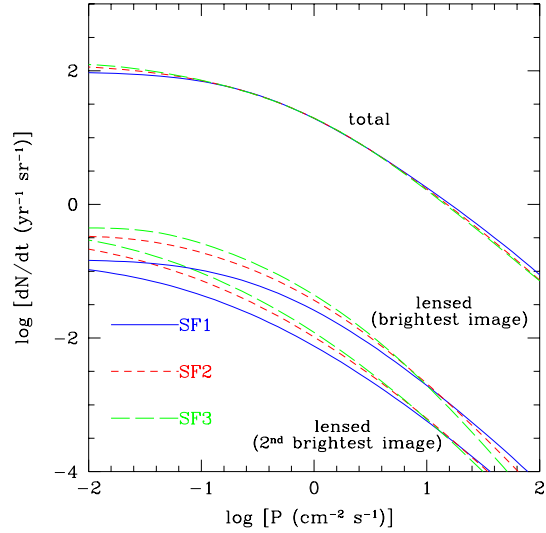}
\caption{Theoretical number counts of GRBs versus apparent peak 
photon flux (assuming a perfect detector with $\epsilon=1$ and full sky 
coverage). The solid, 
short--dashed and long--dashed lines refer to models in which the burst rate
has been assumed to be proportional to the star formation histories SF1, SF2, 
and SF3, respectively. The upper set of curves shows the counts in absence of 
any lensing effect. From top to bottom, the remaining two sets depict
the expected number counts for the brightest and the second brightest 
images of strongly lensed GRBs.}
\label{lensGRB}
\end{figure}

The cumulative rate of lensed GRBs can then be computed as
\be 
{dN\over dt}=\int_0^\infty dz \,\f{dV(z)}{dz} \f{R_{\rm GRB}(z)}{1+z} 
\int_{0}^
{\infty} dL \,\psi(L) \int_{\mu_{\rm min}(L,z)}^{\infty} d\mu \,
\calP(\mu,z) \epsilon[\mu P(L,z)]\;,
\label{counts2}
\ee
where $\mu_{\rm min}(L,z)$ is the minimum 
magnification
needed to detect a source in a flux--limited survey.
The above equation relates 
the number counts to the probability distribution of magnification, 
${\cal P}(\mu,z)$, which is related to $\tau(z)$ as discussed in
Porciani \& Madau (2000).  The detection rates for the 
two brightest detectable images of strongly lensed GRBs are 
compared with the total number counts 
in Figure \ref{lensGRB}. These results have been obtained by considering
the different star formation histories discussed above
(solid, short--dashed, and long--dashed lines refer to SF1, SF2, and
SF3, respectively), and taking $\epsilon(P)=1$. The upper set of curves 
(which lie approximately on top of each other since we impose our models 
to provide the same number counts in the BATSE energy
band) represents the GRB counts in absence of any
lensing effects. 
The $\log N-\log P$ relation has a $-3/2$ slope at the bright end
($P \gta 10 \punits$) and progressively flattens out 
with decreasing limiting flux. The remaining two sets of curves show, 
respectively, 
the count--rates for the brightest and the second brigthest images
of strongly lensed GRBs. Qualitatively, the shapes of these curves
are similar to those of the unlensed counts. The limiting peak flux at  
which they flatten out, however, depends on the assumed star formation
history, and decreases from SF1 to SF3 as the lensing cross--section 
increases with the source redshift.
Note that the rates in Figure \ref{lensGRB} 
are all--sky averages, and the detection 
of the fainter image (the last to reach the observer) {\it does not} imply 
that its brighter counterpart will also be observed. This is because satellite 
experiments generally guarantee only partial sky (and temporal)
coverage and have a rapidly varying field--of--view.
In this case the detection of {\it multiple} images of the
{\it same} event has a much smaller probability.
For a perfect detector -- full sky and time coverage plus $\epsilon=1$ -- 
our models predict the presence of a doubly--imaged event 
every 2557, 1886, and 1615 bursts with $P>1\,\punits$ for SF1, SF2 and SF3, 
respectively. At fainter fluxes, $P>0.01\,\punits$, recurrent events will
be detected every 888, 533, and 420 bursts instead.

Marani (1998) has compared the light curves of the 
brightest $75\%$ events of a sample containing 1235 BATSE bursts.
Her analysis revealed the absence of good lens candidates with
$P>1\,\punits$, a result largely expected because of the low 
efficiency of BATSE at detecting multiply--imaged bursts. As a consequence of 
Earth blockage, BATSE could only monitor $\sim 2/3$ of the sky at 
the same time. Moreover, the trigger was disabled during readout time and 
when the spacecraft was in specific locations.
Depending on declination, the angular exposure 
(i.e. the fraction of time during which burst detection is 
possible in a given direction of the sky) 
of the 4B catalogue ranges between 0.44 and 0.6, with a mean value of 
0.48 (Hakkila \etal 1998). Since the orbital period of {\it CGRO} was 
only $\sim 5000$ s, the BATSE efficiency for 
detecting a burst in a particular direction of the sky varied 
with a characteristic time--scale which was much shorter 
than the typical time--delay between lensed multiple images.
In other words, the phases of the {\it CGRO} orbit at which two mirror
images of a burst could be observed were practically uncorrelated:
the efficiency of multiply--imaged detection is then proportional to 
$\epsilon(\mu_1 P)\times \epsilon(\mu_2 P)$, and roughly 3/4 of lensed 
events remained undetected. We will see in the next section that, because of
the inefficient duty cycle, the probability of detecting a double burst
is quite small even in a more sensitive future experiment like {\it Swift}.  

\section{Discussion}

\begin{figure}
\epsscale{0.4} \plotone{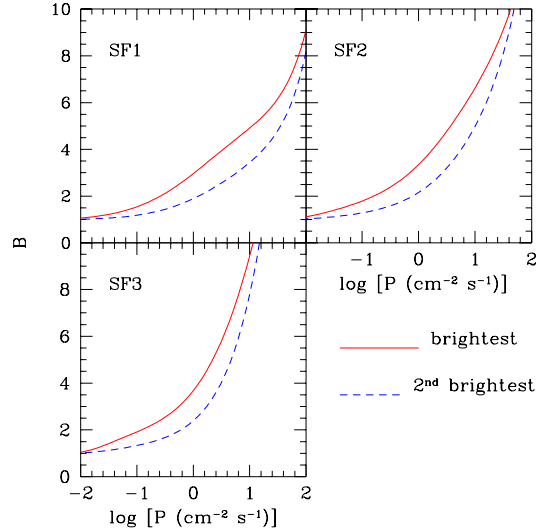}
\caption{The magnification bias $B$ (quantifying the fractional excess
of lensed images in a flux--limited sample of GRBs relative to a volume 
limited sample) is plotted versus limiting peak photon flux.
{\it Solid lines:} Brightest image of strongly lensed GRBs.
{\it Dashed lines:} Second brightest image of multiply--imaged bursts.}
\label{bias}
\end{figure}

In a flux--limited sample, sources that are observed to be gravitationally
lensed include not only the lensed objects that are intrinsically brighter
than the flux limit, but also sources that are intrinsically fainter
but are brought into the sample by the magnification effect
Relative to the optical depth shown in Figure \ref{tau}, lensed systems
will then be overrepresented among GRBs of a given peak flux.
We have quantified this effect from the GRBs number counts, by defining
the magnification bias $B$ as the ratio between the actual flux--limited
counts of lensed GRBs and the counts of lensed events that are intrinsically
brighter than the flux limit,
\be
B(<P)=\f{\displaystyle{
\int_0^\infty
 dz \,\f{dV(z)}{dz} \f{R_{\rm GRB}(z)}{1+z} 
\int_{0}^{\infty} dL\, \psi(L) \,\epsilon
\int_{\mu_{\rm min}(L,z)}^{\infty} {\cal P}(\mu,z) \,d\mu}}
{\displaystyle{
\int_0^\infty
 dz \,\f{dV(z)}{dz} \f{R_{\rm GRB}(z)}{1+z} 
\int_{L(P,z)}^{\infty} dL\, \psi(L) \,\epsilon
\int_{\mu_{\rm min}(L,z)}^{\infty} {\cal P}(\mu,z) \,d\mu}}\;.
\ee
This definition (Porciani \& Madau 2000) generalizes that given in Turner, 
Ostriker \& Gott (1984) by taking into account the redshift
dependence of the lensing optical depth.
The magnification bias for the two brightest images of a GRB 
is plotted in Figure \ref{bias} (for $\epsilon=1$).
At faint fluxes, where lensing becomes a significant effect,  
the bias is small as a consequence of the
flatness of the burst $\log N-\log P$ relation.
Contrary to quasars, where the number of sources at a given flux 
rises steeply at the faint end, there are relatively few GRBs
to be brought into the flux--limited sample by the magnification effect.
The situation is different for bright bursts. In this case,
the steepening of the $\log N-\log P$ relation causes a rapid increase
of the magnification bias. However, as shown in Figure \ref{lensGRB}, 
the absolute rate of lensed events is extremely low for such luminous events.

\begin{figure}
\epsscale{0.4} \plotone{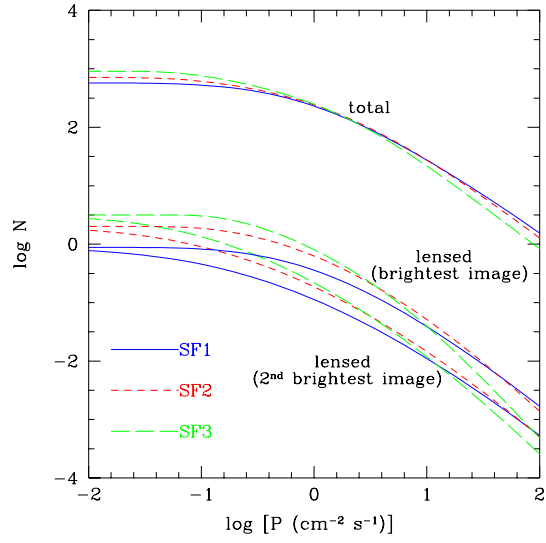}
\caption{Same as Figure \ref{lensGRB}, but for the total counts accumulated
by the BAT onboard {\it Swift} during 3 years of observations.
A field--of--view of 2 sr was assumed.}
\label{swift}
\end{figure}

It is interesting to use our modelling of the GRBs number counts, redshift
distribution, and lensing probability to make predictions for a future 
space mission like the {\it Swift} Gamma Ray Burst Explorer, a 
multiwavelength orbiting observatory selected by NASA for launch in the 2003
\footnote{see http://swift.sonoma.edu/.}.
Its main instrument will be the Burst Alert Telescope (BAT)  
in the 10--150 keV energy range. A X--ray telescope (XRT) and an 
ultraviolet and optical telescope (UVUOT)
complete the onboard instrumentation and will be used to study afterglows
and get accurate determinations of burst locations.
In Figure \ref{swift} we plot our estimates for the GRB number counts 
accumulated by the BAT in a 3--year survey, as a function of limiting 
flux $P$.
These have been computed in the energy band $10-150$ keV, assuming
a field of view of 2 sr and $\epsilon=1$.
The expected sensitivity of the BAT should be close to $0.14\,\punits$ in the 
fully--coded field of view ($0.2\,\punits$ half--coded), assuming 
a flat--topped GRB of duration 2 s and an 8--sigma detection (D. Palmer, 
personal communication). 
Note that, even though the luminosity functions and the overall spatial
density of bursts have been fixed to reproduce the BATSE rates,
the {\it Swift} counts corresponding to SF1 and SF3 disagree by about a factor
of two at both the faint and bright ends. This shows how data from
BATSE and {\it Swift} could be combined to set additional constraints on the
statistical properties of GRBs. 

The number of lensed images that could be detected with the BAT is small. 
As shown in Figure \ref{swift}, in a 3--year
survey with $P>0.1 \, \punits$, {\it Swift} should detect 0.45 
secondary  
lensed images (i.e. the second brightest images of strongly lensed events)
over a total number of 525 bursts for SF1,
0.90 over 611 for SF2, and 1.34 over 760 for SF3.
Even at $P>0.01 \, \punits$, the number of lensed images would not
sensibly increase: 0.78 over 573 for SF1, 1.73 over 715 for SF2, and
2.74 over 908 for SF3. In a SCDM cosmology the numbers are typically 
40\% higher. In any case, since the field of view of BAT covers a small 
fraction of the sky,
the probability of detecting a lensed pair is extremely low.

For particular configurations, lensed pairs could also be detected by combining
BAT and XRT observations. The XRT will make high resolution spectroscopic 
observations of afterglows from the initial acquisition ($\sim 50$ s after 
the burst) for up to 10 hours, while spectrophotometric observations will 
go on for up to 4 days after the burst.
Thus, if the time delay between the different components of a strongly 
lensed system is as small as a few days, the secondary image 
may be easily detected by the XRT during follow--up observations (even when 
too weak to trigger the BAT).  For a given lensing halo, and
at fixed source and lens redshifts, the time delay is anti--correlated with the
magnification: smaller delays correspond to more perfect source--lens 
alignments and thus to larger magnifications.
On the other hand, 
less massive deflectors will always produce relatively short delays.
For example, in our $\Lambda$CDM cosmology, isolated SIS lenses at $z=0.5$, 
deflecting the light emitted by a point source at $z_s=2$ and having
masses of $10^{12}$, $10^{11}$ and $10^{10}$ $M_\odot$, 
will produce maximum delays of 6.6, 1.4 and 0.3 days, respectively.
Thus, the search for short--delayed images of lensed bursts could be optimized
following up those bursts which are located in the vicinity of field 
galaxies having redshifts between 0.5 and 1.
Assuming SIS+NFW lenses distributed in mass according to the PS theory,
and considering a point source at $z_s=2$, we find that $\sim 15\%$ 
of the bursts which generate multiple images will have time delays smaller 
than 4 days. This implies that, on average, only about one lensed pair
should be detected during a 3--year survey.

So far we have confined our attention to axially symmmetric lenses producing
two observable images. Non axisymmetric (e.g. elliptical) potentials, however, 
generally produce 5 images (one of them always strongly demagnified and 
in practice unobservable), and this may increase the odds of a successful
lensing detection. 
As shown by Grossman \& Nowak (1994), the probability of observing 
two or more images out of $n$ above the detection limit
follows a binomial distribution
\be
{\cal P}(\geq 2|n)= \sum_{i=2}^n 
\left({}^n_i \right) p^i (1-p)^{n-i}\;,
\ee
where $p$ is the probability of detecting a single image.
Taking the sky coverage of {\it Swift} ($1/2\pi$) as a representative value
for $p$, one gets: ${\cal P}(\geq 2|2)=0.025$,  ${\cal P}(\geq 2|3)=0.068$ 
and ${\cal P}(\geq 2|4)=0.122$. 
If systems producing 4 sufficiently bright images are common, then
the probability of detecting a recurrent burst is a factor of 5 higher
than estimated by considering axisymmetric lenses only.
For QSOs, doubly--imaged objects account roughly for one--half of all
lensed systems, groups of 4 images contribute for another $30\%$
(see {\it http://cfa--www.harvard.edu/castles/index.htm}). In the case of 
GRBs, the magnitude of this enhancement will depend on their redshift 
distribution. 

\acknowledgments
Support for this work was provided by NASA through ATP grant NAG5--4236 and 
grant AR--06337.10--94A from the Space Telescope Science Institute (P.M.). C.P. 
is supported by a Golda Meir fellowship.

\bigskip\bigskip
\section*{Appendix A. Star formation rate density in different cosmologies}

The estimate for the luminosity density at redshift $z$, obtained by 
combining photometric and spectroscopic data of a galaxy sample, depends 
on the assumed underlying cosmology.
In particular, it comes out proportional to the quantity
$F(z|\Omega_M,\Omega_\Lambda, h_{65})=d_L^2(z)/(dV/dz)$.
Thus,
$$
R_{\rm SF}(z|\Omega_M,\Omega_\Lambda,h_{65})=\f{F(z|\Omega_M,\Omega_\Lambda, h_{65})}
{F(z|1,0,1)} R_{\rm SF}(z|1,0,1)=\f{H(z|\Omega_M,\Omega_\Lambda,h_{65})}
{H(z|1,0,1)} R_{\rm SF}(z|1,0,1)\;, \eqno({\rm A1})
$$
and, expliciting the dependence of the Hubble expansion rate $H$
on the cosmological parameters, we eventually get
$$
R_{\rm SF}(z|\Omega_M,\Omega_\Lambda,h_{65})=
h_{65} \f{[\Omega_M(1+z)^3+\Omega_K(1+z)^2+\Omega_\Lambda]^{1/2}}{(1+z)^{3/2}}
 R_{\rm SF}(z|1,0,1)\;. \eqno({\rm A2}) 
$$

\vfill\eject

\end{document}